# Femtosecond laser nanostructuring in glass with sub-50nm feature sizes

Yang Liao [1], Yinglong Shen [2], Lingling Qiao [1], Danping Chen [2], Ya Cheng [1, *], Koji Sugioka [3] and Katsumi Midorikawa [3]

[1] *State Key Laboratory of High Field Laser Physics, Shanghai Institute of Optics and Fine Mechanics, Chinese Academy of Sciences, P.O. Box 800-211, Shanghai 201800, China*

[2] *Key Laboratory of Material Science and Technology for High Power Lasers, Shanghai Institute of Optics and Fine Mechanics, Chinese Academy of Sciences, P.O. Box 800-211, Shanghai 201800, China*

[3] *Laser Technology Laboratory, RIKEN - Advanced Science Institute, Hirosawa 2-1, Wako, Saitama 351-0198, Japan*

*Electronic mail: ya.cheng@siom.ac.cn*

**Abstract**

We report on controllable production of nanostructures embedded in a porous glass substrate by femtosecond laser direct writing. We show that a hollow nano-void with a lateral size of ~40 nm and an axial size of ~1500 nm can be achieved by manipulating the peak intensity and polarization of the writing laser beam. Our finding enables direct construction of 3D nanofluidics inside glass.

One of the dreams in nano-science and nano-technology is to create three-dimensional (3D) structures of arbitrary geometries and configurations with nanoscale resolution. Femtosecond laser direct writing provides a rapid and extremely flexible approach for achieving this goal [1-2]. So far, by inducing two-photon polymerization (2PP) of photosensitive materials with femtosecond laser pulses, 3D fabrication with a resolution of sub-100 nm has already been achieved by choosing a laser intensity slightly above the threshold value [3]. However, the 2PP technology is intrinsically limited to polymer materials, in which direct formation of thin microfluidic channels cannot be easily achieved due to the difficulty in removing the non-photopolymerized materials contained in the thin channels. On the other hand, although techniques for fabrication in glass have been well established, fabrication of nanostructures with a resolution better than 100 nm in a controllable manner has not been demonstrated [4-6]. In this Letter, we show that by uniquely employing a porous glass as the substrate material, nanogratings consisting of an array of elliptical-shaped hollow nano-voids elongated in the direction of laser propagation can be formed in glass using linearly polarized writing beam. By reducing the peak intensity of the laser pulses close to a threshold value, only a single hollow nano-void in the central area of the focal spot can survive, providing the basic element for construction of 3D nanofluidics inside glass.

In our experiment, home-made high-silicate porous glass samples were used as the substrates, which were produced by removing the borate phase from phase-separated alkali-borosilicate glass in hot acid solution. The composition of the porous glass is approximately 95.5$SiO_2$-4$B_2O_3$-0.5$Na_2O$(wt.%). The pores with a mean size of ~10 nm are homogeneously distributed in the glass and occupy 36% in volume of the glass. Particularly, these pores in the porous glass form a 3D connective network which allows liquid to flow through, enabling direct embedding large scale 3D microfluidic structures in the porous glass by femtosecond laser direct writing [7-8]. A commercially available porous glass with the similar properties as our home-made glass is the VYCOR glass manufactured by Corning Incorporated.

The formation of nanogratings in glass with linearly polarized femtosecond laser pulses has been reported by several groups [9-10]. In the nonporous glass materials, the nanogratings are composed of alternative regions of high and low etching rates [11]. Although the mechanism has not been fully understood, generally this phenomenon can be attributed to a localized near-field enhancement contributed by the plasmon polaritons induced by femtosecond laser irradiation. Interestingly, we found that the nanogratings can also be formed in the porous glass whereas the gratings are composed of an array of hollow nano-voids. To induce the nanograting structure, a high-repetition regeneratively amplified Ti:sapphire laser (Coherent, Inc., Reg 9000) with a pulse duration of ~100 fs, a central wavelength of 800 nm and a repetition rate of 250 kHz was used. The Gaussian laser beam with an initial 8.8 mm diameter was trimmed to1~3 mm-dia. for ensuring a high beam quality, and then was tightly focused into the porous glass immersed in water by a water-immersed microscopy objective (N.A.= 1.10), as shown in Figure 1. The laser power was controlled by using a combination of a polarizer, a waveplate, and a set of neutral density filters. A half-wave plate was used to change the polarization direction of the linearly polarized femtosecond laser. The samples were translated by a computer-controlled XYZ stage with a resolution of 1 μm. In all the experiments, the laser pulses were focused ~200 μm below the surface of samples. To characterize the morphology of the embedded nanograting structures, the samples were cleaved either along the top surface (i. e., XY-plane, Fig. 1) or along the plane perpendicular to the writing direction to access the top or cross section areas of the laser-modified zones. The revealed nanograting structures were directly characterized using a scanning electronic microscopy (SEM, Zeiss Auriga 40). No chemical etching was used before the SEM observations.

Figure 2 presents the top view SEM images of nanograting structures written in porous glass with femtosecond laser pulses linearly polarized in three different directions. The 1D continuous structures were produced by translating the XYZ stage in the horizontal direction. The incident direction and the polarization of the laser beam are indicated in each panel. It can be clearly seen that the nano-gratings are always oriented in perpendicular to the polarization direction of the writing beam, which is consistent with the previous observations in fused silica [12]. However, unlike nonporous glasses, the porous glass chosen for our experiment contains a large number of nanopores. The pores can be either collapsed or deformed under the strong shock wave produced by femtosecond laser induced nano-explosion, providing sufficient room for creation of hollow nano-voids in the explosion regions, as evidenced in Fig. 2. In addition, we have observed that the periodicity of the nanogratings can be varied by controlling the writing conditions, e. g., the writing beam energy and the writing speed. Within a certain range, we found that either higher pulse energy or lower writing speed can lead to smaller periodicities.

We now show that this unique capability of forming hollow nano-gratings in glass can enable controllable 3D nanostructuring inside glass if combined with the threshold effect. Generally speaking, the threshold effect is a highly nonlinear phenomenon which relies on a sharp dependence of the femtosecond-laser-induced photoreaction on the laser peak intensity when the peak intensity is near a threshold value. Such sharp intensity dependence naturally provides a means for achieving a fabrication resolution beyond the diffraction limit if one carefully controls the laser peak intensity in the focal volume so that only in its central region the peak intensity can be higher than the threshold value for inducing a specific photoreaction [13]. However, one significant drawback of superresolution nanofabrication based on the threshold effect is its extreme sensitivity to the peak intensity of the writing laser beam if one would like to push the fabrication resolution to its limit. Therefore, although theoretically there is no limit on the fabrication resolution based on the threshold effect, in reality reliable fabrication resolutions are mostly achieved around 50 nm ~ 100 nm for two-photon polymerization, and sub-100 nm fabrication resolution has never been demonstrated in glass so far due to the different physical mechanism in bulk machining [14].

Figure 3 shows how this difficulty can be overcome. As can been seen in Fig. 2, when the glass sample is translated in the direction perpendicular to the laser polarization, we find that the nanograting structures are oriented along the writing direction. The cross sections of the nanogratings written in this way at different pulse energies (with a same writing speed of 10 μm/s) are shown in Fig. 3. At a relatively high pulse energy of 90 nJ, the number of the elliptical nano-voids in the grating is large, as shown in Fig. 3(a). By reducing the pulse energy to 70 nJ, the number of the nano-voids decreases to only four. It is noteworthy that at this condition, the periodicity of the nanograting apparently increases as evidenced in Fig. 3(b), which also helps reduce the number of nano-voids. Finally, by reducing the pulse energy to 60 nJ, only the central nano-void survives, showing an extremely narrow width of ~37 nm as presented in Fig. 3(c). The evolution of the nanostructures from Fig. 3(a) to 3(c) unambiguously manifests the central role played by the threshold effect behind this phenomenon. In addition, since the formation of the hollow nano-void is enabled by the collapse of the surrounding nano-pores which provide room for the materials to expand within the bulk glass, we can reasonably expect that nano-voids of narrower widths should be achievable by reducing the pore size or the occupation ratio of the pores in the porous glass.

Amazingly, we found that the single nano-voids produced by femtosecond laser pulses can be connected into a continuous structure which enables construction of 3D nanofluidic channels in glass. This is achieved by slowly scanning the focal spot of the femtosecond laser in the porous glass from an open connected to a microchannel or microchamber filled with water at translation speed of 5~10 μm/s . Through the open, the ablation debris can be efficiently driven out from the nanochannel by the bubbles formed in water by the femtosecond laser direct writing. Figure 4a presents a zoom-in view of the cross section of the nanofluidic channel, which shows a slit-like cross section with a width of ~40 nm and a height of ~ 1.5 μm. It can also be seen that the nanochannel is surrounded by small nanopores with sizes ranging from ~5 to ~10 nm. A direct evidence on the formation of a single continuous nanochannel is presented in Fig. 4(b), where a top view of a portion of the nanochannel is shown by its SEM image. Since the height of the channel is only ~1.5 μm, special care has been taken when we polish the glass sample in order to open the top side of the nanochannel. The limited length of the nanochannel in Fig. 4(b) is caused by the fact that the polishing plane is slightly tilted with respect to the horizontal plane where the nanochannel lies in. After the laser writing, we anneal the sample at a high temperature of 1120 °C for 2 h. The postannealing process will collapse all the nano-pores in glass; otherwise the nanochannels are leaky due to the connective networks formed by the nano-pores. To confirm that there are no clogs in the nanofluidic channels, we inject a fluorescent dye solution into the channel and observed that the entire nanofluidic channel has been filled with the dye solution.

More details for producing embedded 3D nanofluidic channels as well as their integration with microfluidic channels have been discussed elsewhere [15].

To summarize, we have demonstrated the direct formation of regularly arranged hollow nanograting structures inside porous glass in a controllable manner using a linearly polarized femtosecond laser. Taking advantage of the threshold effect, the number of the nano-voids in the nanogratings can be controlled by changing the peak intensity of the laser beam. Based on this approach, a single nano-void with a width of 37 nm has been inscribed in glass. Formation of nanofluidic channels has been achieved by connecting the single nano-voids into smooth continuous structures in the porous glass. After a postannealing for collapsing all the pores, the nanofluidic channel can confine and transport fluorescent solutions between two microfluidic channels without leakage and clogging. By further optimization of the pore sizes, 3D nanofluidic channels with even narrower widths could be achieved with our technique, which can be useful tools for applications ranging from nanofluidics research to biomolecular analysis.

**Figure 1** Schematic diagram of inducing nanograting structures inside water-immersed porous glass by femtosecond laser pulses. The sample is translated in the x-y plane, the writing polarization is aligned to the x axis, and the laser incident direction is along the z axis.

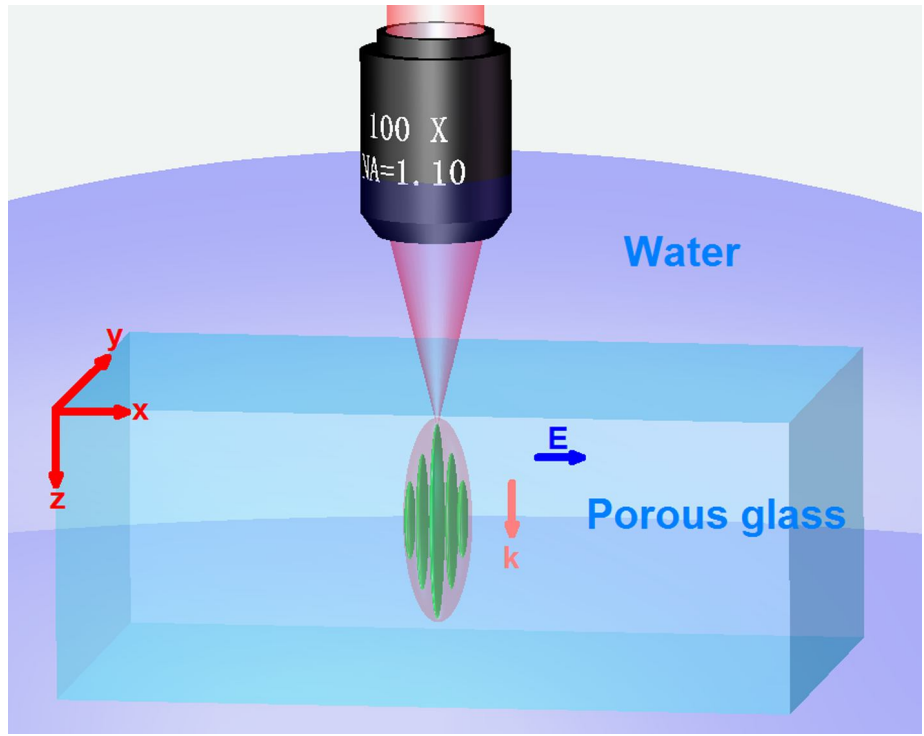

**Figure 2** SEM images of hollow nanograting structures induced by femtosecond laser direct writing with different polarization directions. The laser incident direction (**k**), polarization direction (**E**), and the writing direction (**S**) are indicated in each panel. Images (a) and (b) were obtained with a translation speed of 20 μm/s and pulse energies of 90 nJ and 100 nJ, respectively. Image (c) was obtained with a translation speed of 50 μm/s and the pulse energy 100 nJ..

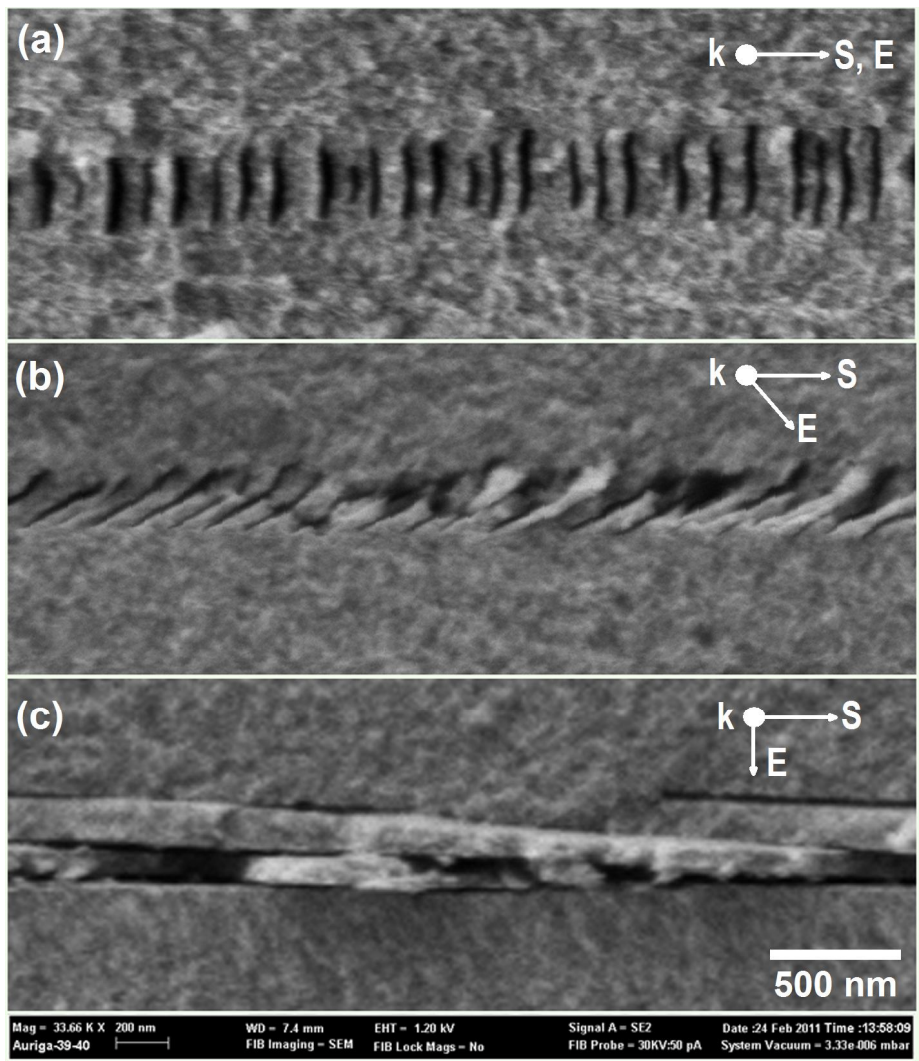

**Figure 3** Evolution from an array of nano-voids to a single nano-void with decreasing laser power. The laser incident direction (**k**), polarization direction (**E**), and the writing direction (**S**) are indicated.

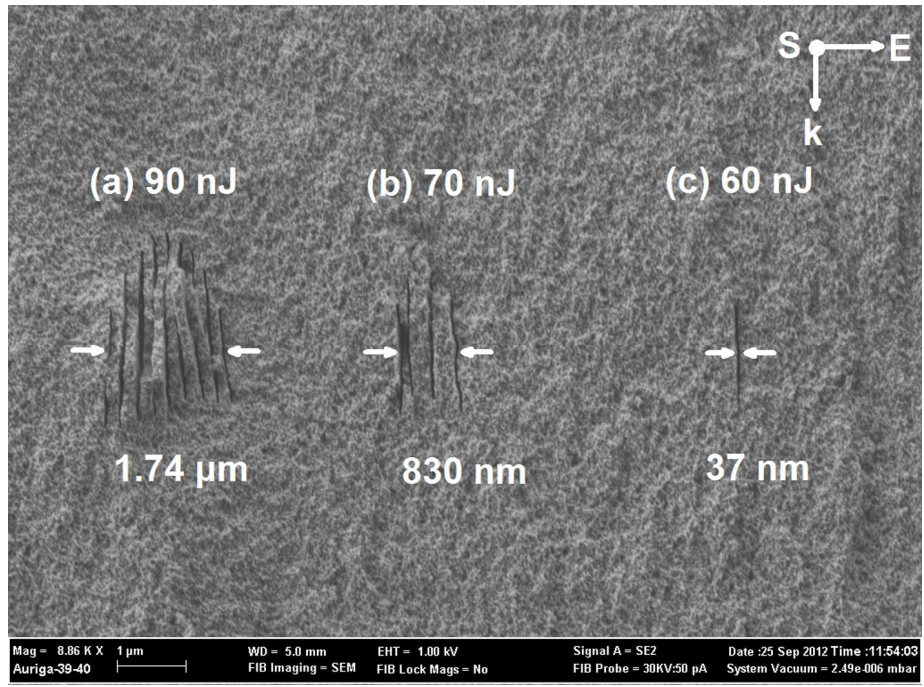

**Figure 4** (a) Cross section view and (b) top view SEM images of a single nanofluidic channel written in the porous glass. (C) Fluorescence microscope image of the postannealed nanofluidic channels filled with fluorescent dyes.

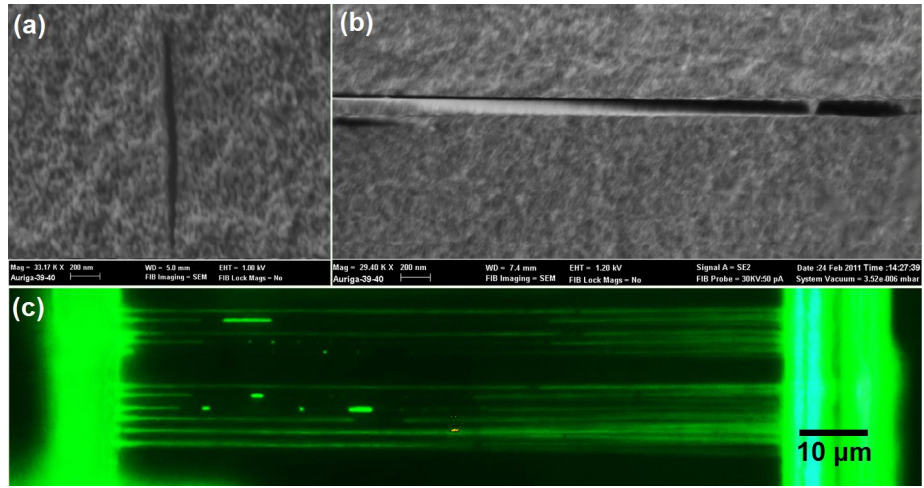